\documentclass[mathleft]{an}
\usepackage[english]{babel}
\usepackage[T1]{fontenc} 
\usepackage{csquotes}
\usepackage{graphicx}
\usepackage{color}
\usepackage{times}
\usepackage{url}
\usepackage{amsmath,amssymb}
\usepackage{natbib}
\bibpunct{(}{)}{;}{a}{}{,}

\newcommand{\tmin}{\ensuremath{\mathrm{min}}}
\newcommand{\tmax}{\ensuremath{\mathrm{max}}}
\newcommand{\efunc}[1]{\ensuremath{\mathrm{e}^{#1}}}
\newcommand{\dif}[1]{\ensuremath{\mathrm{d}#1}}

\begin{document}
\sloppy

% The following seven commands are intended for editorial usage and should
% be ignored by the author(s).
%\Pagespan{}{}% Document's page range. 
% If second parameter is left empty, the last page is computed automatically.
\Yearpublication{2010}%
\Yearsubmission{2010}%
\Month{10}%   
\Volume{331}%  
\Issue{9}% 
\DOI{0.1002/asna.201011425}% 

\title{A New Code for Fourier-Legendre Analysis of Large Datasets -- First
  Results and a Comparison with Ring-Diagram Analysis}
\titlerunning{A new Code for Fourier-Legendre Analysis}

\author{%
  H.-P. Doerr\inst{1}
  \and
  M. Roth\inst{1}
  \and
  A. Zaatri\inst{1}
  \and
  L. Krieger\inst{2}
  \and
  M. J. Thompson\inst{3}%
}
\authorrunning{H.P. Doerr, M. Roth, A. Zaatri, L. Krieger, and M.J. Thompson}
\institute{%
  Kiepenheuer-Institut f\"ur Sonnenphysik,
  Sch\"oneckstra\ss{}e 6,
  79106 Freiburg,
  Germany
  \and
  Institut f\"ur Geophysik,
  Universit\"at Hamburg,
  Bundesstra\ss{}e 55,
  20146 Hamburg,
  Germany
  \and
  School of Mathematics \& Statistics,
  University of Sheffield,
  Sheffield S3 7RH, UK}

\received{2010 Sep 9}
\accepted{2010 Sep 13}
\publonline{2010 Nov 11}


\abstract{%
  Fourier-Legendre decomposition (FLD) of solar Doppler imaging data is a
  promising method to estimate the sub-surface solar meridional flow. FLD is
  sensible to low-degree oscillation modes and thus has the potential to
  probe the deep meridional flow.
  We present a newly developed code to be used for large scale FLD analysis
  of helioseismic data as provided by the Global Oscillation Network Group
  (GONG), the Michelson Doppler Imager (MDI) instrument, and the upcoming
  Helioseismic and Magnetic Imager (HMI) instrument.
  First results obtained with the new code are qualitatively comparable to
  those obtained from ring-diagram analyis of the same time series.
}

\maketitle

\section{Introduction}

From summer 2010 on, the Helioseismic and Magnetic Imager (HMI) instrument
aboard the Solar Dynamics Observatory (SDO) satellite is expected to deliver
continuous high resolution (4\,k $\times$ 4\,k) full-disk Doppler images at
a cadence of 45\,seconds. The availability of this data will surely mark a
new era for helioseismology, but the huge amount of data that HMI will
produce also poses a challenge for helioseismic data processing
pipelines. With the given resolution and cadence, the HMI full-disk Doppler
images alone will sum up to roughly 2~terabytes per month.

The current major sources of solar full-disk Doppler data are the Global
Oscillation Network Group \citep[GONG;][]{Harvey.et.al.1996} and the
Michelson Doppler Imager \citep[MDI;][]{Scherrer.et.al.1995} instrument
aboard the Solar and Heliospheric Observatory (SOHO) spacecraft. They
deliver Dopplergrams with a maximum volume of about 100~gigabytes per month,
each. Both GONG and MDI have been operating for more than one decade and
it is exciting that one can now apply helioseismic techniques to data
spanning about one solar cycle from two independent observatories.

The characteristics of the solar meridional circulation are an important
ingredient to some solar dynamo models, see e.g. the review
by~\citet{dikpati09} and references therein.  A poleward flow of the order
of 10 to 20 m/s can be derived at the surface and in the near-surface layers
by a variety of techniques
~\citep[e.g.][]{duvall79,hathaway96,komm93,woehl01,haber02} but no significant
evidence for a return flow in deeper layers has been detected so far.

Fourier-Legendre spectral decomposition (FLD) is sensitive to low-degree
modes and thus has the potential to detect flows in deep layers. From
encouraging results of previous research \citep{Braun.Fan.1998,
  Krieger.Roth.vdL.2007} and the progress of our ongoing work, we are
confident that the Fourier-Legendre analysis can become a standard tool for
meridional flow measurements below the surface along with ring-diagram
analysis~\citep{hill88} and time-distance helioseismology~\citep{duvall93}.

However, this requires a data processing pipeline that is able to handle
efficiently the available and upcoming helioseismic data. In this paper, we
present a newly developed code for Fourier-Le\-gend\-re analysis that fits
these requirements. The code has been subject to several tests.  Here we
present first results obtained with this code and compare them to results
from ring-dia\-gram analysis of the same time series. It is our intention
making the code available to the helioseismology community as part of the
European Helio- and Asteroseismology Network (HELAS).

\section{Methods}

\subsection{The Fourier-Legendre spectral decomposition}

The ``Fourier-Hankel spectral decomposition'' technique has been used to
study $p$-mode scattering by sun\-spots
\citep{Braun.Duval.Labonte.1987,Braun.et.al.1992,Bogdan.et.al.1993}.
Later the method was also applied for sub-surface meridional flow
measurements by \citet{Braun.Fan.1998} and \citet{Krieger.Roth.vdL.2007}.

In plane geometry, Hankel functions are a good approximation to Legendre
functions and in some of the publications mentioned above they were used
instead of Legendre functions because they are numerically much easier to
compute. In this work we do not use Hankel functions, and to avoid confusion
we prefer to use the more descriptive notation ``Fourier-Legendre
decomposition'' for the remainder of this paper.

Following \citet{Braun.Duval.Labonte.1988} and \citet{Braun.Fan.1998}, the
surface oscillation signal $\delta V(\theta,\,\phi,\,t)$ within an annular
region around a point of interest can be represented as a superposition of
inward and outward traveling waves of the form
\begin{multline}
  %\notag
  \delta V(\theta,\,\phi,\,t) = \sum\limits_{l\,m\,\nu}
  \Bigl[
    A_{lm\nu}X_l^m(\theta)\\
    +
    B_{lm\nu}\left(X_l^m\right)^*(\theta)
  \Bigr]\,
  \efunc{i(m\phi + 2\pi\nu t)},
  \label{eq:surface-wavefield}
\end{multline}
where $A_{lm\nu}$ and $B_{lm\nu }$ are the complex amplitudes of the
oscillation modes of the temporal frequency $\nu$, the harmonic degree $l$
and azimuthal order $m$. The angle $\theta$ equals zero in the center
of the region of interest and increases with radial distance from the
center. The basis functions
\begin{equation}
  X_l^m(\theta) =
  N_l^m \left[P_l^m(\cos\theta) -\frac{2i}{\pi}Q_l^m(\cos\theta)\right]
  \label{eq:Xlm}
\end{equation}
and its complex conjugate $(X_l^m)^*$ are superpositions of the associated
Legendre functions of first and second kind $P_l^m$ and $Q_l^m$, and $N_l^m
= (-l)^m\frac{(l-m)!}{(l+m)!}$ is a normalization factor.

For meridional flow measurements, the center of the annular region is
identical to either the northern or southern pole, so that $A_{lm\nu}$ and
$B_{lm\nu}$ become the poleward and equ\-atorward waves, respectively.

Using equation~(\ref{eq:surface-wavefield}), the mode amplitudes can be
extract\-ed from the measured Doppler signal by 
\begin{multline}
  A_{lm\nu} = 
  \frac{C_l}{2\pi T}
  \int\limits_0^T
  \int\limits_0^{2\pi}
  \int\limits_{\theta_\tmin}^{\theta_\tmax} 
  \!\delta V(\theta,\,\phi,\,t)\,
  \efunc{-i(m\phi + 2\pi\nu t)}\\
  \times (X_l^m)^*(\theta)\, \theta\,
  \dif\theta\, \dif\phi\, \dif t
  \label{eq:extract-Alm}
\end{multline}
and
\begin{multline}
  B_{lm\nu} = 
  \frac{C_l}{2\pi T}
  \int\limits_0^T
  \int\limits_0^{2\pi}
  \int\limits_{\theta_\tmin}^{\theta_\tmax} 
  \!\delta V(\theta,\,\phi,\,t)\,
  \efunc{-i(m\phi + 2\pi\nu t)}\\
  \times X_l^m(\theta)\, \theta\,
  \dif\theta\, \dif\phi\, \dif t
  \label{eq:extract-Blm},
\end{multline}
where $C_l$ is a normalization factor which is approximately given by $\pi
\sqrt{l(l+1)}/[2(\theta_\tmax - \theta_\tmin)]$, $\theta_\tmax$ and
$\theta_\tmin$ mark the latitudinal extent of the annular region, and $T$ is
the total length of the observed time series.

\section{The numerical code}

The decomposition code is written in the C language for efficiency and
portability. It is parallelized using the Message Passing Interface (MPI)
standard. Further steps in the post-processing pipeline such as peak fitting
and the inversion are currently implemented as prototype code in IDL
(Interactive Data Language).

The design goals for the code were:
1) to address the most relevant issues and open questions that arise from
previous publications,
2) to provide the flexibility to easily use data from common sources such as
MDI, GONG and the upcoming HMI,
3) to write a code that is fast enough to process all available data in
reasonable time, and
4) to make the code flexible enough to not only use it for meridional flow
measurements, but also to measure $p$-mode absorption in sunspots as
demonstrated by \citet{Braun.Duval.Labonte.1988}.

We addressed the issues as discussed by \citet{Braun.Fan.1998} and
\citet{Krieger.Roth.vdL.2007}. This includes proper handling of the
spherical geometry by using Legendre functions instead of Hankel functions
(this was already done by \citeauthor{Braun.Fan.1998}, though) and an
inversion for the flow-profile.
Requirement 2) is implemented by following a modular concept: for each data
source, a plug-in module is provided that knows how to read data from the
respective source and forwards the Doppler images to the decomposition
module in a standardized format. Goal 4) will be implemented in the future.

On a modern quad-core processor, one month of GONG data can be processed in
less than one day with a setup as we use it in this paper (see below). The
decomposition into the mode coefficients only takes a minor fraction of the
total runtime and the further steps are not optimized for speed yet, so that
we expect future versions of the code to run much faster. The code also
scales well up to the amounts of data we expect from the HMI instrument
later this year.

\subsection{Data preparation}

The Doppler images are processed in \emph{chunks}. Each chunk consists of a
configurable number of Dopplergrams which usually equals to 60 for the
one-minute cadence of GONG and MDI.

The Dopplergrams are interpolated to an equidistant heliocentric
$\theta,\phi$\,-grid. Effects of the solar rotation and relative movement
between instrument and Sun are removed by subtracting the mean Dopplergram
of one chunk from each Dopplergram in the chunk. Each Dopplergram is then
apod\-ized using a Hann window function to avoid spatial aliasing in the
resulting power spectra. Bad or missing Dopplergrams are detected and the
mode amplitudes of the corresponding timestep are set to zero. A binary mask
is stored along with the time series that allows to properly detect and
account for gaps in the time series.

\subsection{Decomposition \& frequency fitting}

For the decomposition of the mode coefficients, we follow an approach that
was first used by \citet{Brown.1985} for fast Spherical Harmonic
decomposition of solar oscillation data. The integration in longitude in
Eqs.~\eqref{eq:extract-Alm} and~\eqref{eq:extract-Blm} is implemented as a
Fast Fourier Transform (FFT). Compared to a straightforward numerical
evaluation of the integrals, this results in a greatly improved performance
of the overall computation.

Because the associated Legendre functions $P_l^m$ and $Q_l^m$ can only be
calculated reliably using a recursion relation \citep{Zhang.1996}, we
precompute the basis functions $X_l^m$ for all combinations of
$l,\,m$ and $\theta$. Since we are only interested in low azimuthal orders
$m = -25\ldots +25$, the lookup tables consume a few hundred megabytes and
fit easily into main memory.

The resulting time series for the complex amplitudes $A_{lm}$ and $B_{lm}$
are stored as FITS (Flexible Image Transport System) binary tables for the
further steps in the processing pipeline. 
From the power spectra of $A_{lm\nu}$ and $B_{lm\nu}$ the peak frequencies
are determined by fitting asymmetric Lorentzian profiles. From these fits
the frequency differences between the poleward and equatorward propagating
waves are determined.

\subsection{Inversion}

The frequency shift between the poleward- and equatorward propagating waves
is used to invert for the horizontal component of the sub-surface meridional
flow. The frequency shift $\Delta\nu$ is related to the meridional flow by
\citep{Gough.and.Toomre.1983} 
\begin{equation}
  \Delta\nu_{nl} =
  \frac{l \int\limits_0^\infty \langle U_{\rm mer}(r)\rangle K_{nl}(r)\, \dif{r}} 
  {\pi R_\odot \int\limits_0^\infty  K_{nl}(r)\, \dif{r}}\ ,
  \label{eq1}
\end{equation}
where $R_\odot$ is the solar radius, $\langle U_{\rm mer}(r)\rangle$ is the
mean meridional flow averaged over the patch, and $K_{nl}(r)$ are the energy
density kernels, which are calculated from solar 'Model S' by
\citet{Dalsgaard.et.al.1996}. Based on this equation, inversions for the
meridional flow are carried out by employing a SOLA technique (Subtractive
Optimally Localized Averaging) as described in
\citet{Pijpers.and.Thompson.1992,Pijpers.and.Thompson.1994}. Taking
measurement errors of the frequencies into account, the inverted solution
has to be regularized. For simplicity, we used one regularization parameter
for all positions in the Sun. We selected a rather large regularization
parameter in order to give trust only to those modes where a precise
frequency measurement was possible.

\section{Results}

For a first test, we use GONG data from January and February 2006. These
data sets where chosen because at the beginning of this period the GONG duty
cycle was comparably high (91\%) and it is in the middle of the declining
phase of the past solar cycle, so that a minimum of temporal change in the
flow pattern can be estimated.

To estimate the meridional flow as a function of depth and latitude, the
Dopplergrams are subdivided into smaller patches. All patches have a height
of 16~degree in latitude while the longitudinal extent depends on the
position on the disk in order to avoid foreshortened regions. The location
and width of the patches we used is shown in Table~\ref{tab:patches}. These
are the same parameters that were used for positioning 16x16 degree square
patches for the ring-diagram analysis.
\begin{table}
  \caption{Latitudinal position and width of the
    patches in heliocentric coordinates}
  \label{tab:patches}
  \begin{tabular}{rr}\hline
    center latitude [deg] & longitudinal extent [deg]\\ 
    \hline
    $\pm$ 45.0 & -30.0 to +30.0 \\
    $\pm$ 37.5 & -45.0 to +45.0 \\
    $\pm$ 30.0 & -45.0 to +45.0 \\
    $\pm$ 22.5 & -52.5 to +52.5 \\
    $\pm$ 15.0 & -52.5 to +52.5 \\
    $\pm$  7.5 & -52.5 to +52.5 \\
             0 & -52.5 to +52.5 \\
    \hline
  \end{tabular}
\end{table}
For each patch the Fourier-Legendre decomposition is carried out and the
meridional flow is determined from the frequency shifts between pole- and
equatorward propagating waves of harmonic degrees $l = 100$ -- $1000$ and
radial order $n = 0$ -- $11$.

\subsection{Comparison with ring-diagram analysis}

\begin{figure*}
  \centering
  \includegraphics[height=6.54cm,bb=82 372 542 712]{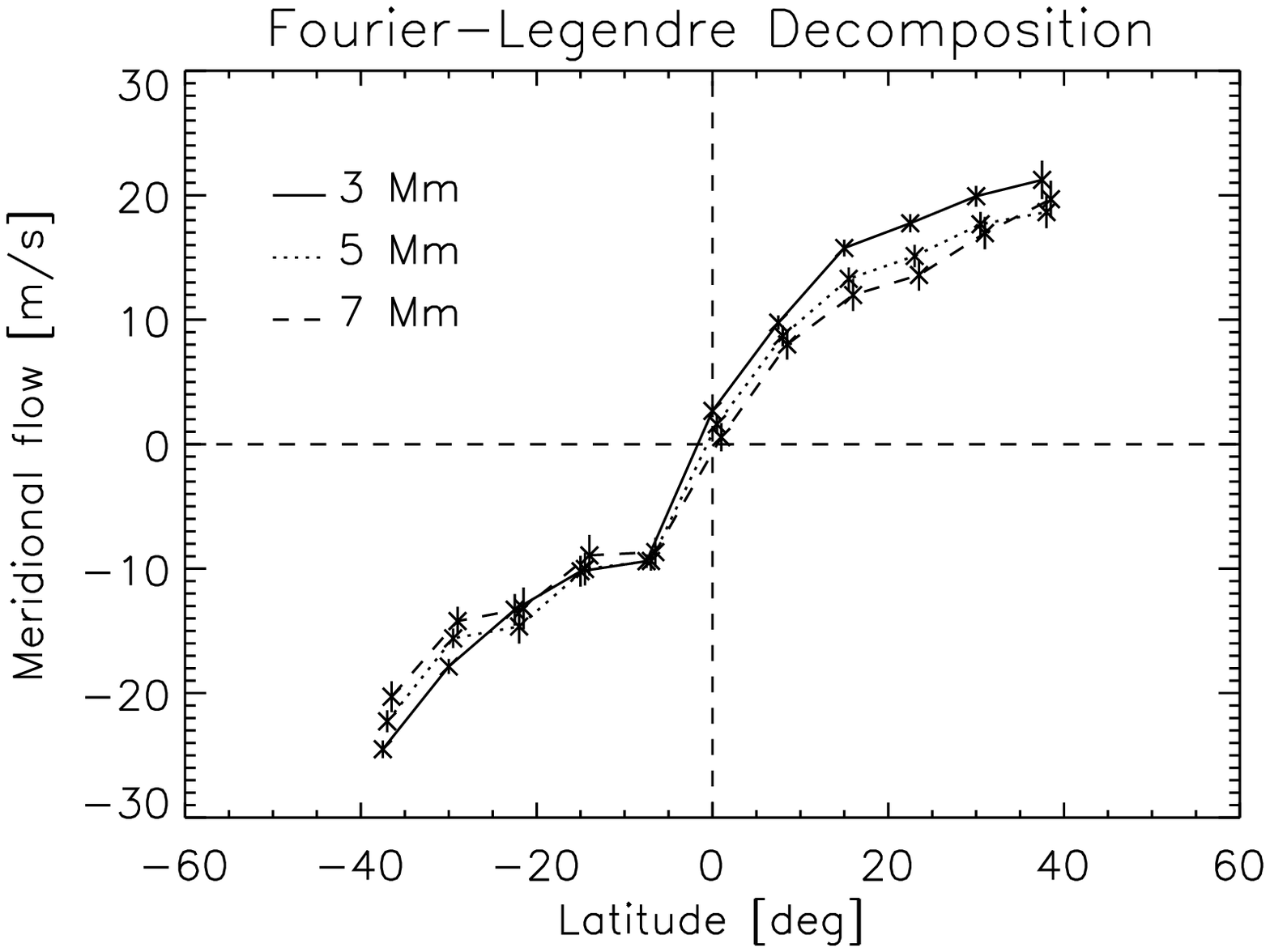}
  \includegraphics[height=6.54cm]{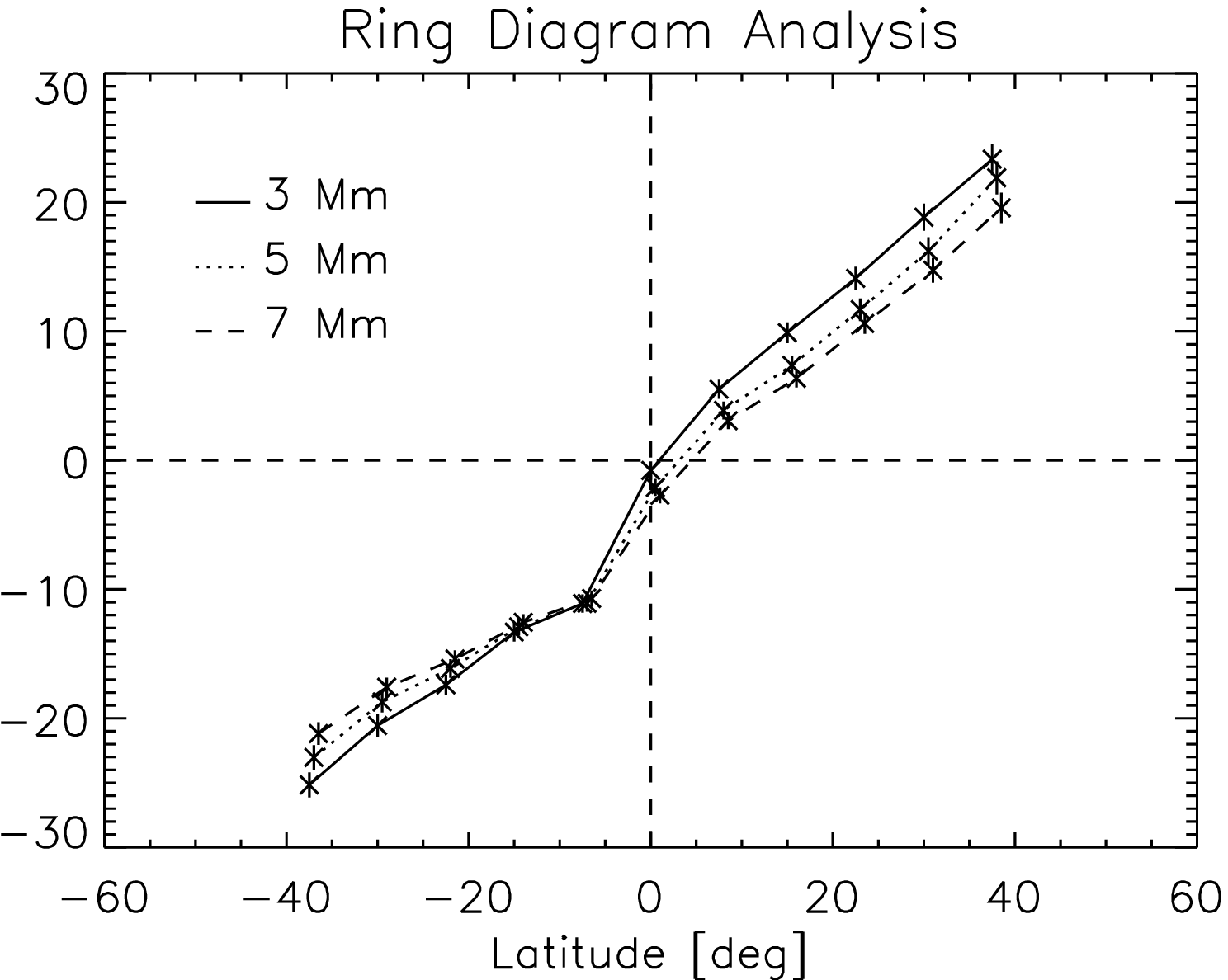}
  \caption{Near surface meridional flow as derived from Fourier-Legendre
    decomposition and ring-diagram analysis of two month of GONG
    data. Errorbars span the $3\,\sigma$ interval at each point. The graphs
    for~5 and~7\,Mm are shifted by 0.5 and 1.0 degrees respectively for
    better visibility.}
  \label{fig:crossect}
\end{figure*}

Figure~\ref{fig:crossect} shows cross sections of the flow profiles at
depths of 3, 5 and 7~Mm obtained from Fourier-Legendre decomposition (left
panel) and ring-diagram analysis (right panel).
With both techniques the measurement error is lowest close to the surface
and at the equator ($\pm 0.2$\thinspace m/s for FLD and ring-diagram
analysis) and highest in the deep layers and at high latitudes ($\pm
0.5$\,m/s for ring-diagram analysis and $0.8$\,m/s for FLD).

Both methods reproduce the same qualitative features of the flow
profile. The direction of the flow is mainly poleward in the order of
20\,m/s. FLD and ring-diagram both favor a weak equator-crossing flow but
with opposite sign. With both techniques the derived flow velocities
increase with latitude. There is a clear asymmetry between both hemispheres
visible in the velocity profiles from both methods. On the northern
hemisphere the velocities tend to decrease with depth, whereas the curves
for the different depth agree within their error marings down to
-22.5~degree latitude. Interstingly, both methods show the same bend in the
velocity profile at -7.5~degree latitude.

\section{Conclusions \& Discussion}

Combined with helioseismic inversion techniques Fourier-Legendre
decomposition of Doppler imaging data can be used to derive the sub-surface
solar meridional flow. The method is sensible to low-degree modes and thus
could greatly increase the range in depth that is currently accessible to
other methods such as ring-diagram analysis.

We developed a new numerical code suitable for Fourier-Legendre analysis of
large sets of Dopplergrams as provided by GONG, MDI and the upcoming HMI
instrument. For a first test of the new code, we compared the near-surface
flow velocities derived from FLD to those obtained from ring-diagram
analysis of the same data sets. Both methods result in qualitatively
comparable flow velocities with small errors in the order of one
percent. However, the absolute values for the velocities obtained with both
methods are not in complete agreement. As for the inversion of the
Fourier-Legendre data one regularization parameter was used for all
positions in the Sun, the discrepancies between the two methods might be due
to this systematic effect.

A careful analysis of possible sources of further systematic errors needs to
be carried out in future. For the case of FLD this also includes the impact
of leakage between modes of neighboring degree $l$ and order $m$ which is
not taken into account in the present paper. Possible solutions are either
to use only a limited set of values of $l$ for which orthogonality of the
Legendre functions is guaranteed \citep{Braun.Duval.Labonte.1988}. Another
possibility is to include the covariance matrix of the mode coefficients in
the procedure for the frequency determination of the modes.

Inversions for the meridional flow in deeper layers of the Sun as well as
long-term studies of the variability of the flow will be carried out when
the open issues are resolved in the near future.

\acknowledgements
The authors thank D. Braun, A. Kosovichev, and H. Schunker for useful
discussions on Fourier-Hankel decomposition and mode fitting.
The authors acknowledge support from the European Helio-
and Asteroseismology Network (HELAS) which is funded as Coordination Action
by the European Commission's Sixth Framework Programme.
This work utilizes data obtained by the Glo\-bal Oscillation Network Group
(GONG) program, managed by the National Solar Observatory, which is operated
by AURA, Inc. under a cooperative agreement with the National Science
Foundation. The data were acquired by instruments operated by the Big Bear
Solar Observatory, High Altitude Observatory, Learmonth Solar Observatory,
Udaipur Solar Observatory, Instituto de Astrof\'{\i}sica de Canarias, and
Cerro Tololo Interamerican Observatory.

%\newpage%%%%%%%%%%%%%%%%%%%%%%%%%%%%%%%%%%%%%%%%%%%%%%%%%%%%%%

%\bibliographystyle{plainnat}
\bibliographystyle{apj}
\bibliography{references}

\end{document}